# Ferroelasticity-Controlled Ultrafast Spin Dynamics in 2D Altermagnetic $V_2STeO$

Tianxia Guo[1,2], Yinlu Gao[3,4], Ying Zhao[5], Zhipeng Hou[6*], Jijun Zhao[1,2], Xue Jiang[1,2*]

[1] *Guangdong Basic Research Center of Excellence for Structure and Fundamental Interactions of Matter, Guangdong Provincial Key Laboratory of Quantum Engineering and Quantum Materials, School of Physics, South China Normal University, Guangzhou 510006, China*

[2] *Quantum Science Center of Guangdong-Hong Kong-Macao Greater Bay Area, Shenzhen-Hong Kong International Science and Technology Park, Shenzhen 518000, China*

[3] *School of Physics and Electronic Engineering, Jiangsu University, Zhenjiang, Jiangsu 212013, China*

[4] *Jiangsu Engineering Research Center on Quantum Perception and Intelligent Detection of Agricultural Information, Jiangsu University, Zhenjiang 212013, China*

[5] *School of Electronic Engineering, Huainan Normal University, 232038, China*

[6] *Guangdong Provincial Key Laboratory of Optical Information Materials and Technology, Institute for Advanced Materials, South China Academy of Advanced Optoelectronics, South China Normal University, Guangzhou, China*

## Abstract

Ultrafast demagnetization induced by femtosecond optical excitation provides a fundamental route toward high-speed magnetic information processing. However, achieving nonvolatile and programmable control of nonequilibrium spin response remains a major challenge. Here, using first-principles calculations and real-time time-dependent density functional theory (rt-TDDFT), we investigate polarization-dependent ultrafast spin dynamics in two-dimensional ferroelastic altermagnetic $V_2STeO$. We show that ferroelastic switching not only generates distinct spin-current responses but also strongly modifies the ultrafast demagnetization behavior through reversible reconstruction of the low energy electronic structure. In particular, the ferroelastic distortion reshapes the symmetry-protected momentum-dependent spin splitting and orbital anisotropy associated with altermagnetism, thereby altering the polarization-dependent optical excitation channels and the resulting angular momentum transfer pathways. Consequently, ferroelastic states exhibit qualitatively distinct demagnetization dynamics, ranging from rapid nonthermal quenching to nonmonotonic evolution accompanied by transient recovery. These pathways can be selectively activated by the combined action of ferroelastic switching and light polarization. Our results reveal a fundamental interplay among ferroelastic symmetry, altermagnetic spin splitting, nonequilibrium carrier dynamics, and ultrafast angular-momentum transfer, and establish ferroelastic switching as a symmetry-engineering strategy for programmable ultrafast spin manipulation in two-dimensional altermagnetic spintronic devices.

## I. Introduction

Understanding and controlling the nonequilibrium dynamics of quantum materials has become a central goal of condensed-matter physics, motivated by both fundamental interest and the prospect of ultrafast functionalities beyond equilibrium limits. Ultrafast demagnetization induced by femtosecond optical excitation provides a unique window into angular-momentum transfer in magnetic materials driven far from equilibrium and underpins emerging ultrafast spintronic technologies[1–3]. A key challenge is to understand how angular momentum is transferred and redistributed among coupled charge, spin, orbital, and lattice degrees of freedom on femtosecond timescales[4]. This challenge is particularly compelling in two-dimensional (2D) magnets, where reduced dimensionality, enhanced electronic correlations, pronounced orbital anisotropy, and crystal-symmetry constraints give rise to complex nonequilibrium spin dynamics that are highly sensitive to the polarization of optical excitation[5–9].

Several microscopic pathways for ultrafast angular-momentum transfer have been proposed in 2D magnetic systems. In $FePS_3$ nanoflake, the disappearance of a 3.2 THz coherent phonon mode above the Néel temperature ($T_N$) revealed spin-lattice coupling as an efficient channel for angular-momentum transfer[10]. Laser-driven antiferromagnetic to ferrimagnetic switching has been predicted in 2D $MnPS_3$, where spin-selective charge transfer mediates ultrafast angular-momentum redistribution on a timescale of tens of femtoseconds[11]. In $CrI_3/WSe_2$ heterojunctions, femtosecond spin injection transiently magnetizes the $WSe_2$ layer, highlighting interfacial spin-transfer as another mechanism for ultrafast spin manipulation[12]. Despite these advances, a unified microscopic picture remains elusive because the relative importance of competing angular-momentum transfer pathways depends sensitively on crystal symmetry, orbital anisotropy, and the resulting optical selection rules. Consequently, establishing a reversible and nonvolatile strategy for controlling ultrafast spin dynamics through symmetry engineering remains an outstanding challenge.

Recently proposed 2D altermagnets (AM) offer a unique opportunity to address this challenge because their momentum-dependent spin splitting originates from crystal symmetry rather than conventional spin–orbit coupling (SOC)[13-19]. As a result, their spin texture, optical selection rules, and nonequilibrium spin dynamics are intrinsically governed by crystal symmetry and can therefore be manipulated through symmetry reconstruction. Ferroelastic switching provides a natural route to such control by reversibly and nonvolatile modifying the in-plane lattice symmetry and crystallographic orientation[20]. Previous studies have demonstrated that ferroelasticity effectively tunes orbital anisotropy, electronic transport, optical transitions, and magnetic anisotropy in 2D ferroelastic semiconductors $Nb_2SiTe_4$ and $Nb_2GeTe_4$, as well as magnetic CrSX (X = Cl, Br, I)[21,22]. In ferroelastic altermagnets, however, symmetry reconstruction extends beyond conventional property tuning by directly reshaping symmetry-protected momentum-dependent spin splitting and modifying polarization-selective optical excitation pathways. Therefore, ferroelastic altermagnets provide a unique platform for linking real-space symmetry engineering with momentum-space spin dynamics.

In this work, we identify 2D $V_2STeO$ as a prototypical ferroelastic altermagnet and reveal how ferroelastic symmetry governs ultrafast spin dynamics under polarization-resolved femtosecond optical excitation. Using first-principles calculations combined with real-time time-dependent density-functional theory (rt-TDDFT), we show that ferroelastic switching profoundly modifies spin-current generation and the competition between different ultrafast demagnetization channels by modulating orbital anisotropy and polarization-dependent optical selection rule. Consequently, the dominant angular-momentum transfer pathway can be selectively controlled through the interplay between ferroelastic symmetry switching and pump polarization. Our results establish ferroelastic symmetry engineering as a general strategy for manipulating ultrafast angular-momentum transfer in low-dimensional altermagnets, opening new opportunities for symmetry-controlled ultrafast spintronics.

## II. Methods

Density-functional theory (DFT) calculations were performed using the projector-augmented-wave (PAW) method as implemented in the Vienna ab initio simulation package (VASP)[23,24]. The exchange-correlation interaction was described within the generalized-gradient approximation (GGA) of Perdew-Burke-Ernzerhof (PBE) functional[23,25]. A plane-wave cutoff of 500 eV and a Monkhorst-Pack k-point mesh of 15×15×1 was employed for structural optimization and electronic-structure calculations. A vacuum spacing of at least 20 Å was introduced along the out-of-plane direction to eliminate spurious interlayer interactions between periodic images. The atomic structures were fully relaxed until the total energy and atomic forces converged to within $10^{-5}$ eV and 0.01 eV/Å, respectively. The optimized lattice constants of the $V_2STeO$ monolayer are a = b = 4.07 Å. The real-time time-dependent density-functional theory (rt-TDDFT) is performed using the ELK code[26, 27]. An in-plane linearly polarized laser pulse was applied as the external driven field. A 9×9×1 k-point grid was employed to sample the Brillouin zone, and the time step was set to 4.8 as. To describe the strong correlated correction of V 3*d* electrons, the DFT + U method was employed with an effective Hubbard parameter of $U_{eff} = 4$ eV, consistent with previous studies[28].

## III. Results and Discussion

To investigate how ferroelastic symmetry engineering governs ultrafast angular-momentum transfer in altermagnets, a material platform combining ferroelasticity with symmetry-protected momentum-dependent spin splitting is required. Monolayer $V_2STeO$ naturally fulfills these requirements and therefore serves as an ideal model system. Its crystal structure is illustrated in **Figure 1(a)**. Starting from the square-lattice $V_2(Se, Te)_2O$ compound with the *P*4/*mmm* space group, monolayer $V_2STeO$ can be viewed as a noncentrosymmetric Janus derivative obtained by replacing the two equivalent out-of-plane anion layers with inequivalent S and Te atoms. This structural asymmetry breaks inversion and out-of-plane mirror symmetries while preserving the essential in-plane symmetry operations of the square V sublattice.

We first determine the magnetic ground state of monolayer $V_2STeO$. Total energy calculations identify the Néel antiferromagnetic (AFM) configuration as the lowest energy state (**Figure S1**[29]), stabilized predominantly by O-mediated V-O-V super-exchange interactions. With the Néel vector oriented along the $y$ ($b$) direction, the magnetic structure preserves only the $m_{xz}$, $m_{yz}T$, and $C_{2z}T$ symmetries, while the fourfold rotation $C_{4z}$ and diagonal mirror symmetries are broken. Combined with the absence of both PT and $\tau$T symmetries, this magnetic symmetry permits momentum-dependent spin splitting, a defining characteristic of altermagnetism. The corresponding electronic band structure is shown in **Figure 1(b)**. Monolayer $V_2STeO$ is an AM semiconductor with an indirect band gap of 0.28 eV. The electronic states around the X and Y valleys exhibit opposite spin polarizations localized predominantly on the $V_1$ and $V_2$ sublattices, whereas the bands along the Γ-M direction remain spin degenerate because of symmetry protection. This alternating spin polarization between the X and Y valleys is a key fingerprint of altermagnetism and directly reflects the underlying symmetry-protected momentum-dependent spin splitting.

To elucidate the microscopic origin of the altermagnetic spin splitting, we further analyze the orbital-projected band structure. The band edge states are dominated by the $d_{yz}$ and $d_{xz}$ orbitals of the two V sublattices. Owing to the symmetry of the square lattice, $V_1$-$d_{yz}$($d_{xz}$) and $V_2$-$d_{xz}$($d_{yz}$) orbitals remain energetically degenerate. Remarkably, their dominant orbital characters are interchanged between the X and Y valleys, giving rise to momentum-dependent orbital reconfiguration, as illustrated in **Figure 1(c)** and **Figure S2**[29]. This valley-dependent orbital alternation provides the microscopic origin of the symmetry-protected momentum-dependent spin splitting that characterizes altermagnetism. In addition, the projected density of states (PDOS) shows non-negligible hybridization between the V-$d$ and S/Te-$p$ orbitals near the band edges, whereas the O-$p$ states contribute only weakly in the same energy range, consistent with the orbital-projected band structure shown in **Figure S2**[29].

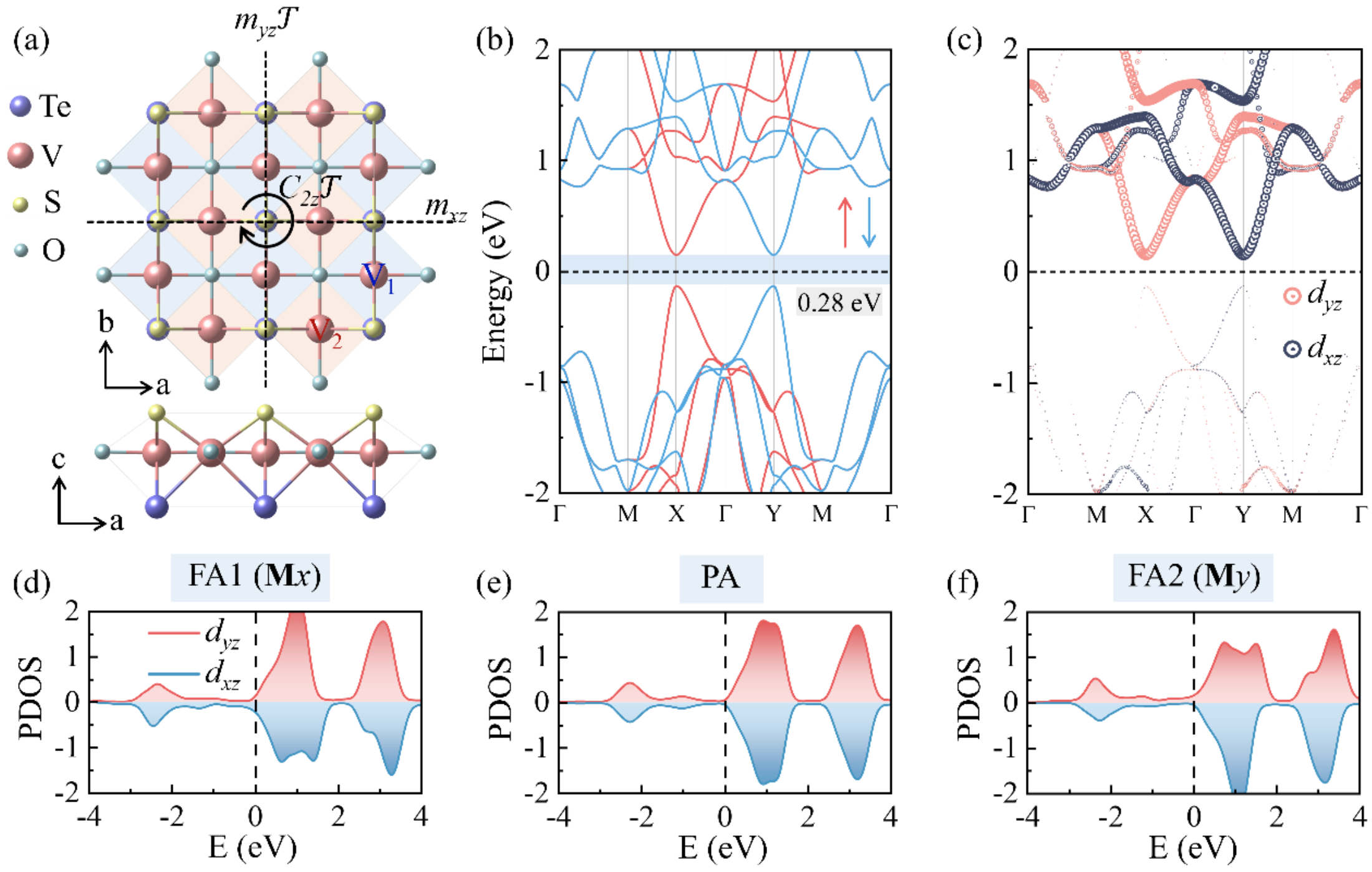

**Figure 1**. (a) Top and side views of the crystal structure of monolayer $V_2STeO$. V atoms with opposite spin orientations are highlighted in blue and pink. (b) Spin-resolved band structure and (c) *d*-orbital-projected band structure of monolayer $V_2STeO$. Projected density of states (PDOS) of the (d) FA1 ($a < b$), (e) PA, and (f) FA2 ($a > b$) phases. The optimized lattice constants of the ferroelastic variants are ($a$, $b$) = (3.87, 4.14) Å for FA1 and (4.14, 3.87) Å for FA2.

Having established the intrinsic AM electronic structure of monolayer $V_2STeO$, we next examine how ferroelastic switching modifies its orbital configuration and magnetic anisotropy. The switching pathway between the two rectangular ferroelastic phases (FA1 and FA2) was obtained by continuously tuning the uniaxial strain along the *a* axis, as illustrated in **Figure S3**[29]. The calculated energy profile shows that the square paraelastic (PA) phase is the ground state, while the AM order remains robust throughout the FA switching process (**Table S1**[29]). A key consequence of ferroelastic distortion is the lifting of the $d_{xz}/d_{yz}$ orbital degeneracy characteristic of the PA phase. As revealed by the orbital-resolved PDOS in **Figure 1(d-f)**, FA switching induces a pronounced, momentum-dependent reconstruction of the low-energy electronic states (**Figure S4** and **S5**[29]). In the FA1 state (a < b), the spin-up $d_{yz}$-derived states around the X valley remain gapped, whereas the symmetry-related spin-down $d_{xz}$-derived states around the Y valley develop a Dirac crossing that passes through the Fermi level, giving rise to a spin-selective conducting channel. In the opposite FA2 state (a > b), the low-energy electronic topology is reversed. The Dirac crossing shifts from the Y valley to the X valley, while the previously gapless branch becomes semiconducting. Consequently, ferroelastic switching reversibly transfers the spin-selective conduction channel between symmetry-related momentum valleys.

The magnetic anisotropy also depends strongly on the ferroelastic state (**Table S2**[29]). Whereas the PA phase exhibits an in-plane easy magnetic plane with nearly degenerate Néel-vector orientations along the $x$ and $y$ directions, the FA1 phase favors the magnetic easy axis of $x$ direction ($M_x$), while the FA2 phase switches the easy axis to the $y$ direction ($M_y$). Therefore, ferroelastic switching simultaneously controls the $d_{yz}/d_{xz}$ orbital splitting, the location of the low energy spin selective Dirac conducting channel, and magnetic anisotropy. More importantly, ferroelastic switching does not simply transfer the conducting channel between two symmetry-related altermagnetic states. By reconstructing the momentum-dependent electronic structure, ferroelastic switching fundamentally reshapes the polarization-dependent optical selection rules, thereby providing the electronic basis for the distinct light-induced spin dynamics discussed below.

To establish a reference for understanding the influence of ferroelastic switching, we first investigate the ultrafast spin dynamics of the intrinsic PA phase under linearly polarized femtosecond excitation along $x$ and $y$ directions. The Néel vector is initialized along the $y$ direction, such that the ground-state magnetization is dominated by the $M_y$ component on both $V_1$ and $V_2$ sublattices. As shown in **Figure 2(a)** and **2(b)**, the dominant magnetic component decreases rapidly following optical excitation for both pump polarizations. The two magnetic sublattices exhibit nearly indistinguishable temporal evolutions, indicating the highly symmetric spin response of the square PA phase. Although the overall demagnetization dynamics are qualitatively similar for the two polarization directions, a clear polarization dependence is still observed, implying that the underlying angular-momentum transfer processes remain sensitive to the polarization-selective optical excitation pathways. To clarify the microscopic origin of this behavior, we next focus on the *y* oriented Néel-vector as a representative configuration, which enables a direct comparison between laser excitation parallel and perpendicular to the magnetic order without implying an intrinsic energetic preference for this direction.

Under *x*-polarized excitation [**Figure 2(a)**], the $M_y$ component of the V-*d* magnetic moment decreases rapidly following photoexcitation, indicating an ultrafast destabilization of the AM order. During the first 10 to 15 fs, $M_y$ is reduced by approximately 1.09 $\mu_B$ and 0.85 $\mu_B$ on the $V_1$ and $V_2$ sublattices, respectively. This initial demagnetization is followed by a transient nonmonotonic regime between approximately 25 and 45 fs, during which $M_y$ undergoes a partial recovery before entering a second stage of demagnetization. At later times, finite transverse $M_x$ components emerge on both V sublattices, reaching approximately 0.15 and 0.22 $\mu_B$ for $V_1$ and $V_2$ sublattices, respectively. These results demonstrate that the photoinduced spin response is not limited to a longitudinal reduction of the magnetic moment but also involves ultrafast spin reorientation.

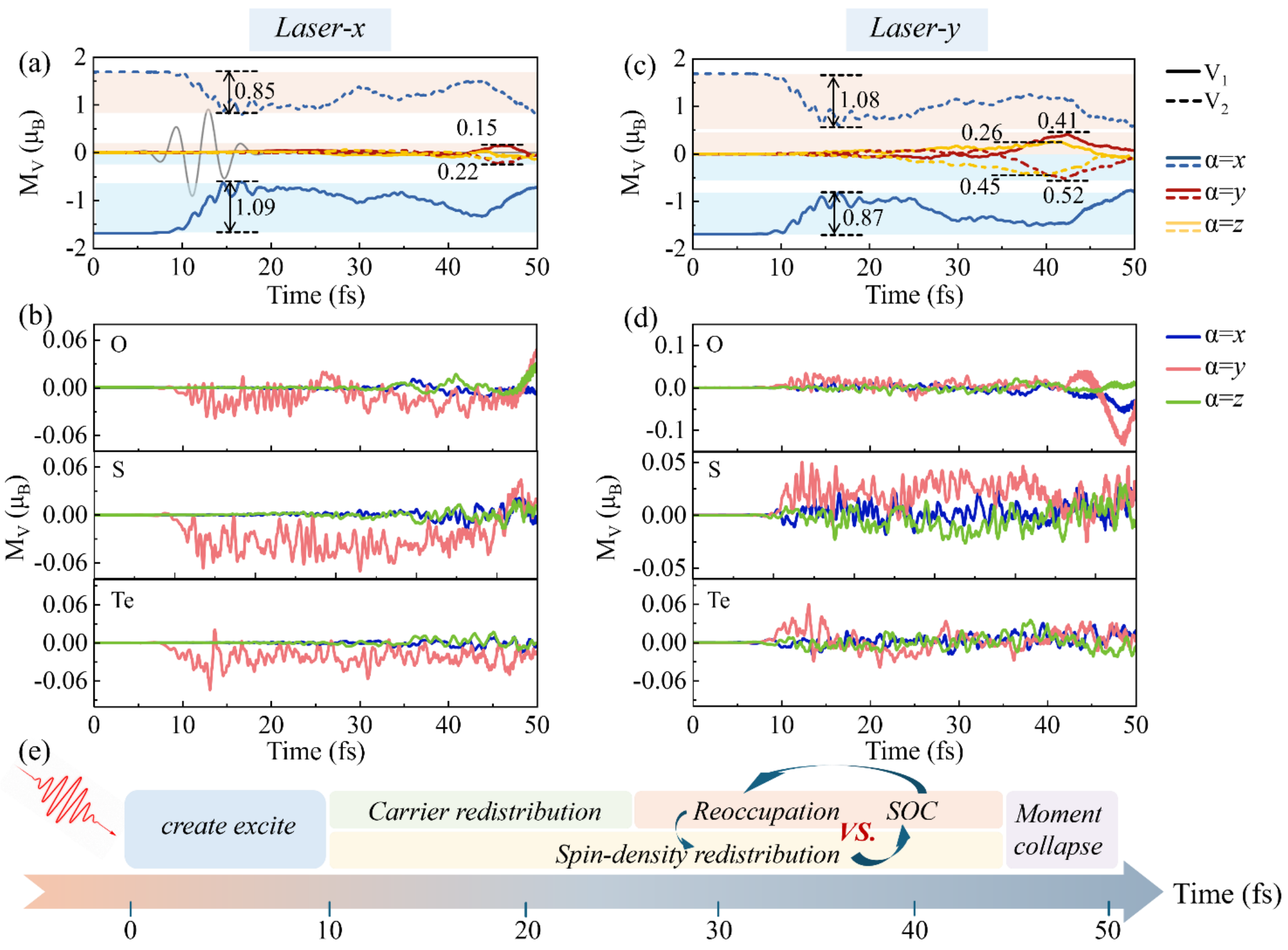

**Figure 2**. Time evolution of the $x$-, $y$-, and $z$-components of the magnetic moments ($M_x$, $M_y$, and $M_z$) on the V, O, S, and Te atoms in monolayer $V_2STeO$ under linearly polarized femtosecond laser excitation in the PA states: (a, b) $x$-polarized excitation and (c, d) $y$-polarized excitation. Solid and dotted curves in (a) and (c) represent the magnetic responses of the $V_1$ and $V_2$ sublattices, respectively. (e) Schematic illustration of the competing electronic processes governing ultrafast demagnetization following photoexcitation, including photoexcitation, nonequilibrium carrier excitation and electronic reoccupation, charge transfer, spin density redistribution, spin-current generation, and SOC-mediated spin reorientation.

The observed sublattice-asymmetric demagnetization originates from the combined effects of symmetry breaking and polarization-selective optical excitation. under $x$-polarized excitation, the $m_{yz}T$ symmetry relating the two magnetic sublattices is no longer preserved, allowing spin-split V-$d$ states with distinct orbital and sublattice characters to be excited differently. The transient recovery of $M_y$ reflects the nonequilibrium redistribution of spin density within the V–O/S/Te network, accompanied by charge transfer between the V-*d* orbitals and the ligand *p* states. As the nonequilibrium carrier population evolves, spin–orbit coupling generates an effective torque that gradually rotates the local magnetic moments away from the initial easy-axis direction, giving rise to the emergence and subsequent growth of the transverse $M_x$ component. A similar coexistence of sublattice-asymmetric demagnetization and spin canting has recently been predicted in the 2D AM $Fe_2WTe_4$, where laser-induced nonequilibrium carrier populations likewise drive simultaneous longitudinal demagnetization and noncollinear spin rotation[30].

Meanwhile, the ligand atoms (O, S, and Te), despite carrying negligible magnetic moments in the ground state, also develop pronounced photoinduced magnetic moments, as shown in **Figure 2(b)**. This observation demonstrates that the ultrafast magnetic response extends beyond the reduction of local V-*d* moments and involves a redistribution of spin-density throughout the V-O/S/Te network. After ~35 fs, finite transverse $M_x$ and $M_z$ components emerge on all ligand sites, revealing that the photoinduced dynamics evolve beyond purely longitudinal demagnetization into a regime of cooperative spin reorientation involving both the magnetic V atoms and their surrounding ligands. Among the ligand atoms, the induced $M_y$ oscillations are significantly stronger on the S and Te sites, reaching amplitudes of up to 0.06 $\mu_B$, whereas the corresponding response on O remains below 0.03 $\mu_B$. This contrast originates from the orbital character of the states near the Fermi level, which are dominated by V-*d* and S/Te-*p* orbitals, while the contribution from O-*p* states comparatively weak. Consequently, S and Te participate more actively in the photoinduced carrier redistribution and the associated angular-momentum transfer. Moreover, the stronger spin–orbit coupling of the heavier Te atom further facilitates the generation of spin components transverse to the initial Néel vector direction. The simultaneous emergence of transverse spin components on both V and ligand atoms therefore reflects a cooperative spin reorientation process driven by the interplay between nonequilibrium carrier dynamics and SOC, as schematically illustrated in **Figure 2(e)**.

Under $y$-polarized excitation, as shown in **Figure 2(c)**, the early-time magnetic response closely resembles that observed under $x$-polarized excitation, indicating that the primary carrier-driven demagnetization channel remains essentially unchanged. The dominant $M_y$ component on both $V_1$ and $V_2$ decreases rapidly after photoexcitation, with a magnitude comparable to that under $x$-polarized excitation. The key difference emerges during the subsequent evolution, where substantially larger transverse spin components develop. The induced $M_x$ ($M_z$) reach ~ 0.41 $\mu_B$ (0.26 $\mu_B$) and ~ 0.52 $\mu_B$ (0.45 $\mu_B$) on $V_1$ and $V_2$, respectively, at around 43 fs, significantly exceeding those obtained under $x$-polarized excitation. These results demonstrate that $y$-polarized excitation more effectively promotes ultrafast spin reorientation while leaving the initial longitudinal demagnetization pathway largely unaffected. The enhanced transverse response originates from the distinct symmetry constraints associated with the pump polarization. Under $y$-polarized excitation, the $m_{xz}$ symmetry is broken, allowing SOC-mediated transverse torques to develop more efficiently. The nonequilibrium carrier population generated by optical excitation therefore produces a stronger effective torque perpendicular to the initial Néel vector, driving the local moments away from the initial $y$ direction toward the $x$ and $z$ directions. Consequently, the role of light polarization is not to modify the primary demagnetization mechanism itself, but rather to regulate the onset and magnitude of the subsequent spin reorientation.

A similar trend is observed for the ligand atoms. Under $y$-polarized excitation, O, S, and Te all exhibit enhanced transverse magnetic responses compared with those under $x$-polarized excitation. Among them, the O atom displays the most pronounced enhancement after ~ 43 fs, with the induced $M_y$ component exceeding 0.11 $\mu_B$ and the transverse $M_x$ component reaching ~ 0.05 $\mu_B$, as shown in **Figure 2(d)**. In contrast, the magnetic responses on S and Te remain comparatively uniform throughout the evolution. Although O contributes only weakly to the band edge electronic states involved in the initial optical excitation, its central role in the V-O-V superexchange pathway allows it to participate more effectively in the subsequent redistribution of spin density and angular momentum. The enhanced magnetic response of O therefore indicates that the photoinduced spin dynamics progressively extend from the initially excited V-*d* and S/Te-*p* states into the entire V-O-S/Te coordination network through nonequilibrium rehybridization.

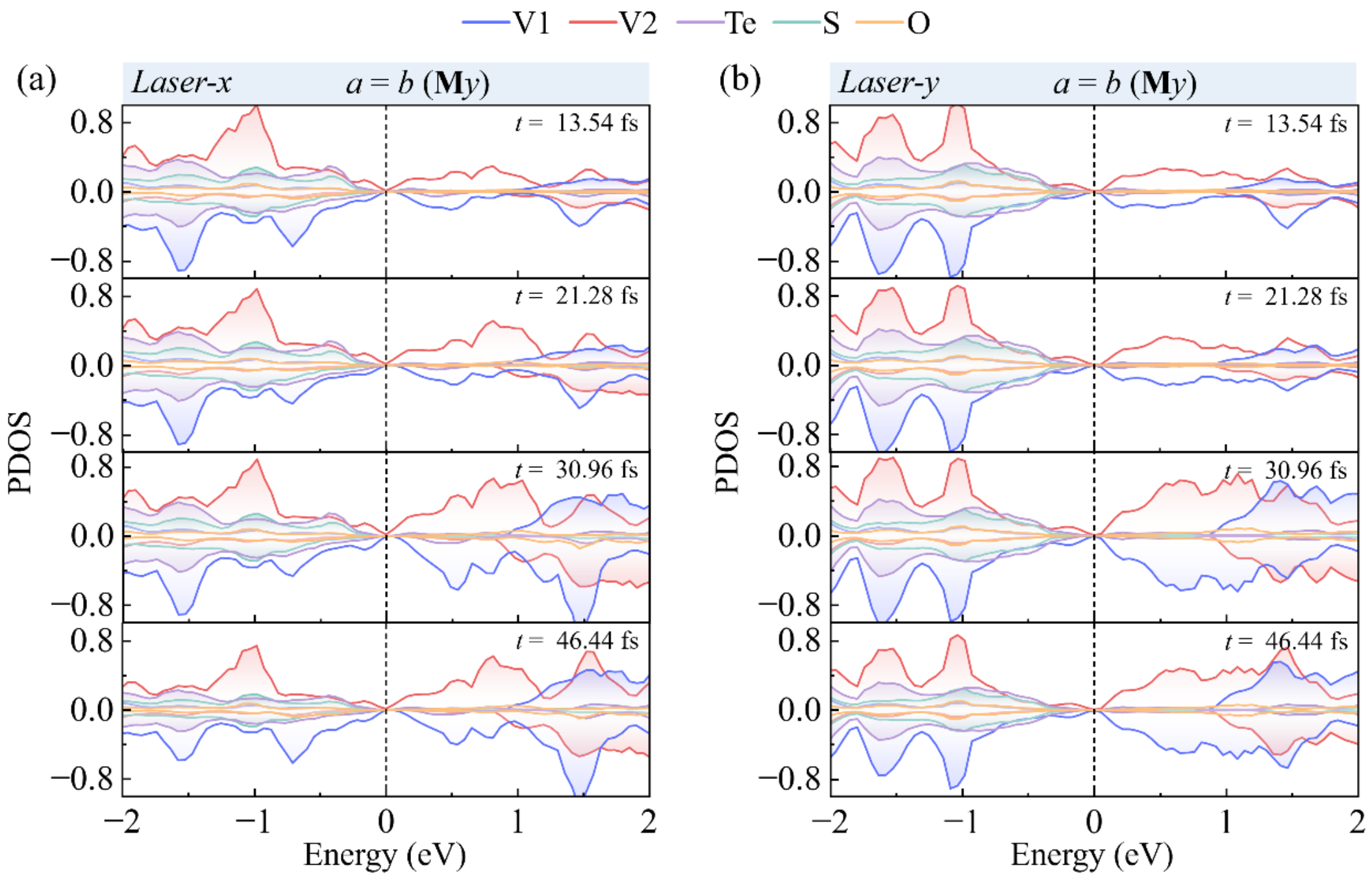

**Figure 3**. Time evolution of the spin-resolved projected density of states (PDOS) of monolayer $V_2STeO$ in the PA state under (a) *x*- and (b) *y*-polarized laser excitation. The PDOS is shown at selected delay times following photoexcitation. Blue and red curves represent the contributions from the $V_1$ and $V_2$ sublattices, respectively.

To further elucidate the microscopic origin of the polarization-dependent spin dynamics, we analyze the time evolution of the spin-resolved PDOS, as shown in **Figure 3**. For both pump polarization, the photoinduced electronic structure changes are concentrated primarily within the low energy states around the Fermi level. During the first few tens of femtoseconds, optical excitation promotes electrons from occupied valence states into low lying conduction states, resulting in a pronounced redistribution of spin-resolved occupations. The associated nonequilibrium carrier redistribution reduces the spin polarization of the V-derived states near the Fermi level, providing the microscopic origin of the rapid longitudinal demagnetization under both polarization directions. The key difference emerges during the subsequent evolution of the low energy electronic structure. Under *x* polarized excitation, the spin-resolved PDOS (**Figure 3**) exhibits a pronounced sublattice asymmetry between $V_1$ and $V_2$ lattices. In contrast, the photoinduced electronic responses of the two sublattices remain considerably more similar under *y*-polarized excitation throughout the nonequilibrium evolution. This distinction originates from the different symmetry constraints imposed by the pump polarization. For *y*-polarized excitation, the optical perturbation largely preserves the $m_{yz}T$ symmetry, maintaining nearly balanced low-energy excitation channels on the two V sublattices. By contrast, *x*-polarized excitation breaks this symmetry, lifting the equivalence between $V_1$ and $V_2$ sublattices and allowing low-energy electronic occupations to evolve independently, thereby producing a progressively stronger sublattice asymmetry. Importantly, the enhanced electronic asymmetry induced by *x*-polarized excitation does not directly translate into stronger spin reorientation. Instead, it is primarily reflected in the unequal redistribution of electronic occupations on the two V sublattices. In contrast, *y*-polarized excitation couples more effectively to SOC-mediated transverse spin torques, leading to a substantially larger buildup of the $M_x$ and $M_z$ components and consequently, stronger ultrafast spin reorientation. These results demonstrate that the essential distinction between the two pump polarizations lies not in the initiation of longitudinal demagnetization, but in the symmetry-controlled evolution of the nonequilibrium electronic structure, which determines the efficiency of transverse angular-momentum transfer.

Overall, the ultrafast spin dynamics of the intrinsic PA phase can be understood as a two-stage process. The initial stage is governed by polarization-independent longitudinal demagnetization driven by nonequilibrium carrier redistribution within the low-energy hybridized states around the Fermi level. The subsequent evolution is dictated by the symmetry-controlled nonequilibrium electronic structure, which determines the efficiency of spin–orbit-coupling-mediated transverse spin torques and, consequently, the extent of ultrafast spin reorientation. This intrinsic dynamical response establishes the benchmark for understanding how ferroelastic symmetry reconstruction modifies angular-momentum transfer in the ferroelastic phases discussed below.

Having established the intrinsic ultrafast spin dynamics of the high-symmetry PA phase, we next investigate how ferroelastic symmetry reconstruction modifies the photoinduced angular-momentum transfer. As demonstrated above, ferroelastic distortion lifts the degeneracy of the low-energy $d_{xz}$ and $d_{yz}$ orbitals through anisotropic *p*-*d* hybridization, thereby reconstructing the polarization-selective excitation channels. Consequently, the nonequilibrium spin dynamics become strongly anisotropic and can be effectively controlled by the pump polarization. For the FA1 state, the magnetic easy axis lies along the *x* direction. Following $x$-polarized photoexcitation, the V moments undergo a rapid demagnetization of the longitudinal $M_x$ component, as shown in **Figure 4(a)**. During the first 10-18 fs, direct optical excitation dominates the magnetic response, reducing $M_x$ by ~ 1.12 and 0.89 $\mu_B$ on the $V_1$ and $V_2$ sublattices, respectively. transient recovery accompanied by weak oscillations before entering a second demagnetization stage at ~ 38 fs, during which $M_x$ decreases further toward complete moment collapse. A qualitatively similar nonmonotonic evolution was also observed in the PA phase, indicating that the transient recovery is an intrinsic feature of the nonequilibrium electronic dynamics rather than a consequence of ferroelasticity itself. Similar multistage ultrafast demagnetization has also been reported theoretically and experimentally in 2D $Fe_3GeTe_2$[31, 32]. In contrast, the entire nonmonotonic evolution in $V_2STeO$ is completed within approximately 50 fs, substantially faster than the phonon-assisted dynamics reported for 2D $Fe_3GeTe_2$. This ultrashort timescale indicates that the transient recovery originates predominantly from competing nonequilibrium electronic occupation channels rather than lattice-mediated relaxation. Meanwhile, SOC-induced spin mixing continuously accumulates throughout the evolution, gradually driving the system beyond purely longitudinal demagnetization toward transverse spin reorientation.

The ligand atoms exhibit a markedly different magnetic response from that of the PA phase. owing to the ferroelastic distortion, the S and Te already possess finite $M_x$ components in the ground state, whereas the O atom remains nearly nonmagnetic. Following photoexcitation [**Figure 4(b)**], the $M_x$ components of S, Te, and O exhibit sustained oscillations but show little enhancement during the initial rapid demagnetization of the V moments. This behavior indicates that the early-time dynamics remain largely confined to the V-*d*-dominated hybridized states near the Fermi level, while the ligand atoms response mainly through induced spin polarization. Only after ~ 38 fs, when the V moments enter the second demagnetization stage, O, S, and Te become increasingly involved in the redistribution of spin density through the combined effects of SOC and V-O/S/Te hybridization. Their magnetic moment increases substantially, accompanied by the gradual emergence of transverse $M_y$ and $M_z$ components.

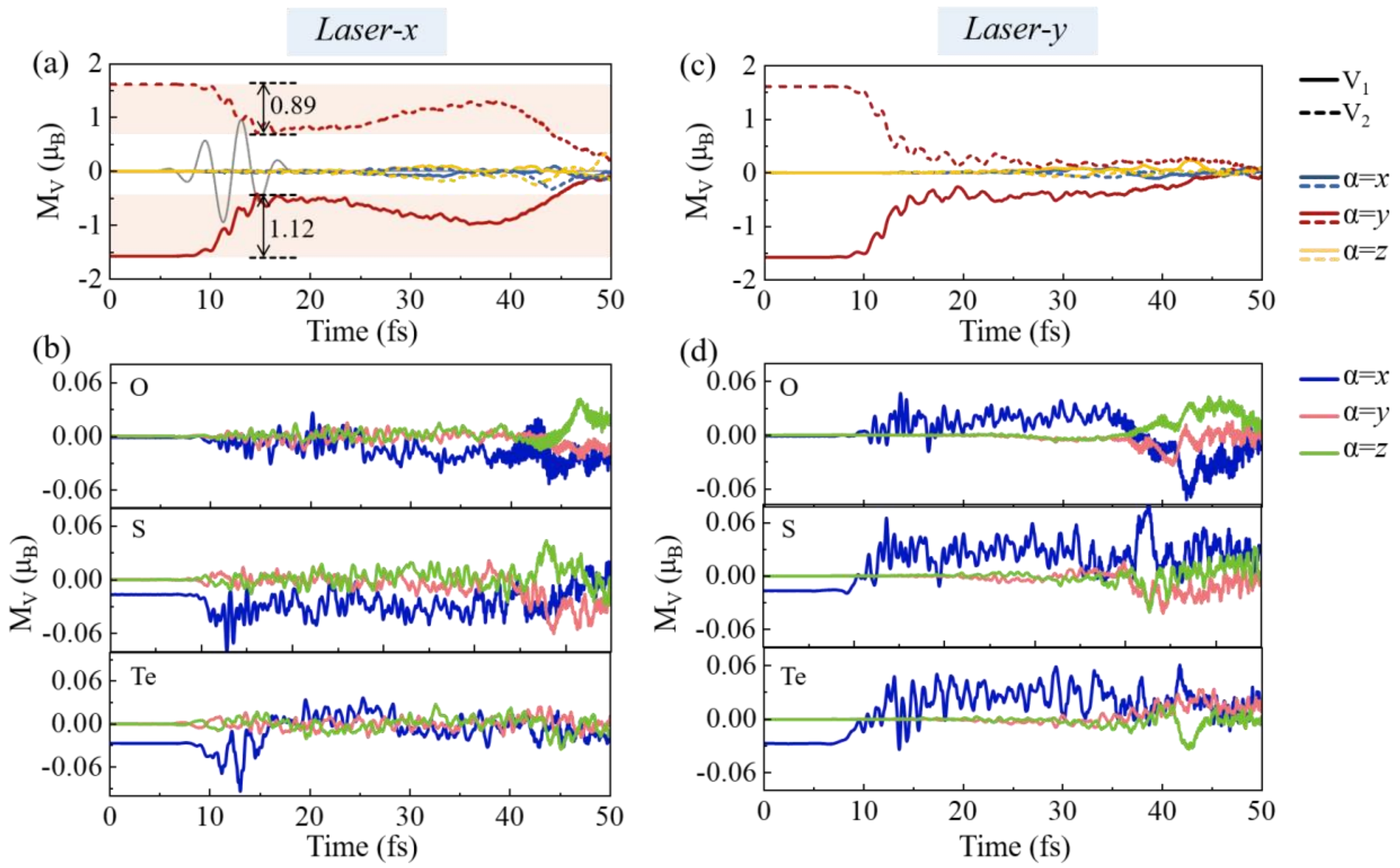

**Figure 4**. Time evolution of the *x*-, *y*-, and *z*-components of the magnetic moments on the V, O, S, and Te atoms in the FA1 ferroelastic phase of monolayer $V_2$STeO under (a, b) *x*- and (c, d) *y*-polarized femtosecond laser excitation. Solid and dotted curves in (a) and (c) denote the magnetic moment of the $V_1$ and $V_2$ sublattice, respectively.

Under *y*-polarized excitation, the FA1 state exhibits a markedly different magnetic response from that observed under *x*-polarized excitation. As shown in **Figure 4(c)**, the longitudinal $M_x$ component on the V sites collapses rapidly and approaches zero within approximately 10-18 fs, indicating a faster and more complete suppression of the initial magnetic moment. In contrast to *x*-polarized case, no evident transient recovery of the longitudinal moment is observed during the subsequent evolution. Instead, the magnetic moment continues to decrease nearly monotonically, resulting in a deeper nonthermal demagnetization. This pronounced polarization dependence cannot be attributed solely to the relative orientation between the pump polarization and the initial magnetization direction, since the same polarization geometry in the high-symmetry PA phase produces only modest differences in the spin dynamics. The underlying origin is the ferroelastic reconstruction of the low-energy electronic structure. The anisotropic *p*-*d* hybridization lifts the orbital degeneracy near the Fermi level and reconfigures the polarization-selective excitation channels, allowing *y*-polarized excitation to couple more efficiently to the low-energy spin-polarized states that stabilize the longitudinal $M_x$ component. As a result, the electronic recovery channel identified under *x*-polarized excitation is strongly suppressed, resulting in a continuous depletion of the longitudinal spin polarization throughout the nonequilibrium evolution. Despite the substantially enhanced demagnetization of the V sites, the ligand atoms exhibit magnetic responses qualitatively similar to those under *x*-polarized excitation, as illustrated in **Figure 4(d)**. The induced magnetic moments on S and Te remain considerably stronger than those on O, indicating that the post-demagnetization redistribution of spin density continues to proceed predominantly through the V–S/Te hybridization network, whereas the involvement of O remains comparatively delayed. This behavior suggests that ferroelasticity primarily modifies the nonequilibrium electronic response of the magnetic V sublattices, while the ligand atoms participate progressively in the subsequent redistribution of spin density and angular momentum.

To elucidate the electronic origin of polarization-dependent demagnetization dynamics in the FA1, we analyze the time evolution of the spin-resolved PDOS at representative delay times, as shown in **Figure 5**. Under *x*-polarized excitation [**Figure 5(a)**], the valence-band spectral weight is substantially depleted at 13.54 fs and 21.28 fs, accompanied by a corresponding increase in the conduction-band occupation. This behavior reflects rapid photoinduced charge redistribution across the Fermi level and accounts for the initial collapse of the longitudinal magnetic moment. At 30.96 fs, the V-derived spectral weight around 1 eV above the Fermi level becomes significantly enhanced, indicating a partial reoccupation of the low-energy spin polarized states. This transient restoration of spectral weight establishes an electronic recovery channel that temporarily replenishes the states supporting the longitudinal $M_x$ component. By 46.44 fs, the recovered spectral weight is largely depleted as carriers continue to populate higher-energy conduction states, leading to the closure of the electronic recovery channel and the onset of the second stage of moment collapse. By contrast, under *y*-polarized excitation [**Figure 5(b)**], the valence-band spectral weight decreases continuously while the conduction-band occupation increases throughout the nonequilibrium evolution, exhibiting a larger overall carrier redistribution without the intermediate-time spectral recovery observed under *x*-polarized excitation. Consequently, the low-energy spin-polarized states near the Fermi level remain persistently depleted, and the electronic recovery channel is effectively absent. Without the transient reoccupation of the states supporting the longitudinal $M_x$ component, the magnetic moment undergoes a nearly monotonic collapse, resulting in a faster and deeper nonthermal demagnetization. These results demonstrate that the pronounced polarization dependence of the FA1 state originates from the ferroelastic reconstruction of the low-energy electronic structure rather than from the optical field itself. By reconfiguring the polarization-selective excitation pathways, ferroelastic symmetry breaking governs the availability of the electronic recovery channel and thereby controls the efficiency of angular-momentum transfer and the anisotropy of the ultrafast spin dynamics.

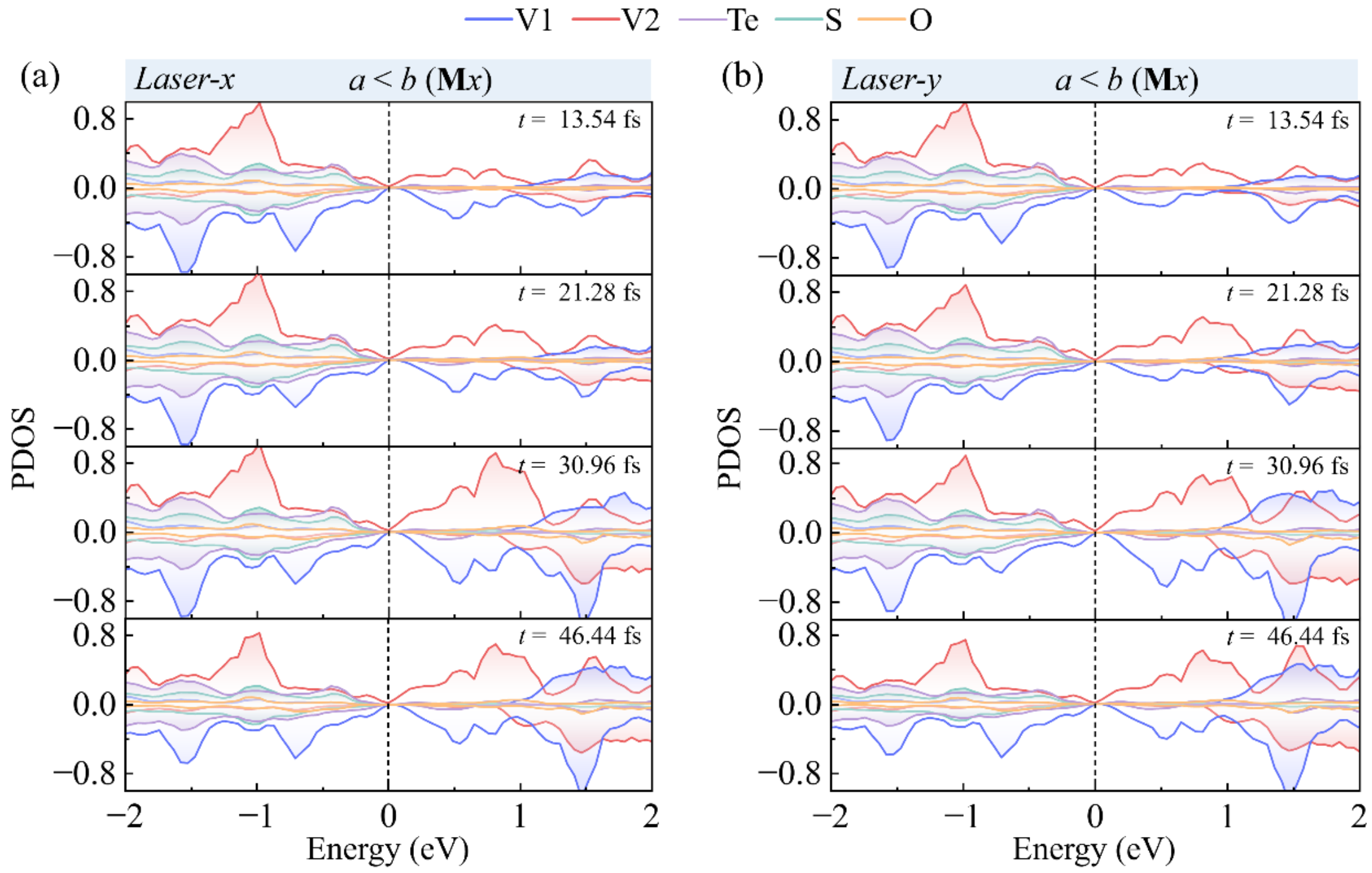

**Figure 5**. Time evolution of the spin-projected density of states (PDOS) of monolayer $V_2STeO$ in the FA1 state at selected delay times (13.54, 21.28, 30.96, and 46.44 fs) under (a) *x*- and (b) *y*-polarized laser excitation. The colored curves represent the PDOS contributions from $V_1$, $V_2$, Te, S, and O, as indicated in the legend.

The FA2 state is the ferroelastic counterpart of FA1 obtained by interchanging the lattice constants *a* and *b*, and exhibits a nearly mirrored ultrafast spin response, as shown in **Figure S6**[29]. Specifically, the rapid monotonic demagnetization observed under *y*-polarized excitation in FA1 is transferred to *x*-polarized excitation in FA2, whereas the nonmonotonic demagnetization accompanied by a transient electronic recovery channel is shifted from *x* polarization in FA1 to *y* polarization in FA2. A corresponding interchange is also observed in the spin-resolved PDOS (**Figure S7**[29]), confirming that the polarization-selective electronic recovery channel is reversibly switched between the two ferroelastic variants. These results demonstrate that ferroelastic switching not only reorients the magnetic easy axis but also reversibly reconstructs the low-energy electronic structure that governs photoinduced angular-momentum transfer. Consequently, the polarization dependence of the ultrafast spin dynamics can be reassigned in a reversible and nonvolatile manner through ferroelastic symmetry engineering. The ferroelastic pair FA1 and FA2 therefore establishes a direct correspondence between real-space symmetry switching and nonequilibrium spin dynamics, providing a general strategy for programming ultrafast demagnetization pathways in low-dimensional altermagnets.

Rather than involving an abrupt transfer of spin angular momentum from the V atoms to the surrounding ligands, the photoinduced magnetic response evolves through a cooperative redistribution of spin density and angular momentum across the entire V–O/S/Te network. As the nonequilibrium electronic occupations evolve, both the V atoms and the surrounding ligands participate progressively in this redistribution, ultimately giving rise to a multicomponent magnetic state accompanied by ultrafast spin reorientation. Ferroelastic symmetry switching reshapes this cooperative process by reconstructing the low-energy excitation pathways responsible for photoinduced angular-momentum transfer, thereby enabling reversible and nonvolatile control of polarization-dependent ultrafast spin dynamics.

## IV. Conclusion

In summary, we have systematically investigated the polarization-dependent ultrafast spin dynamics of the 2D ferroelastic altermagnet $V_2STeO$ using first-principles calculations and real-time time-dependent density-functional theory. $V_2STeO$ hosts two switchable ferroelastic variants that reversibly reconfigure the spin-polarized electronic structure near the Fermi level and generate distinct polarization-dependent excitation channels. Based on this ferroelastic tunability, we reveal a strong interplay among crystal symmetry, optical polarization, and ultrafast angular-momentum transfer. In the high-symmetry PA phase, both *x*- and *y*-polarized excitations induce similar longitudinal demagnetization at early times, while producing different SOC-mediated transverse spin responses during the subsequent evolution due to their different symmetry constraints. In contrast, the ferroelastic states exhibit pronounced polarization selective ultrafast spin dynamics. When the optical polarization is aligned with the magnetic easy axis, $V_2STeO$ exhibits a nonmonotonic evolution involving initial demagnetization, a transient electronic recovery channel, and renewed demagnetization. When the polarization is perpendicular to the easy axis, the low-energy spin-polarized states supporting the longitudinal magnetic moment are continuously depleted, resulting in faster and deeper nonthermal demagnetization. These contrasting behaviors originate from ferroelasticity-driven reconstruction of orbital anisotropy and polarization-dependent low-energy excitation pathways. By integrating ferroelasticity, altermagnetism, and femtosecond spin dynamics within a single $V_2STeO$ platform, our work establishes ferroelastic symmetry engineering as an effective strategy for achieving nonvolatile and polarization-selective control of nonequilibrium angular-momentum transfer and ultrafast spin responses in low-dimensional altermagnets.

## Acknowledgments

This work was supported by the National Natural Science Foundation of China (Grants No. 12274050, and No. 12534012), and Guangdong Provincial Quantum Science Strategic Initiative (Grant No. GDZX2401002).